\documentclass{aa}
\usepackage{amssymb}
\usepackage{graphics}
\usepackage{here}
\usepackage[dvips]{epsfig}
\usepackage{natbib}
\bibpunct{(}{)}{;}{a}{}{,}

\begin{document}     

\title{Statistical methods of automatic spectral classification and
  their application to the Hamburg/ESO Survey}

\author {N. Christlieb\inst{1}
\and L. Wisotzki\inst{2}
\and G. Gra\ss{}hoff\inst{3}  
}

\institute{Hamburger Sternwarte, Universit\"at Hamburg, Gojenbergsweg 112,
   D-21029 Hamburg, Germany\\
   \email{nchristlieb@hs.uni-hamburg.de}
\and Institut f\"ur Physik, Universit\"at Potsdam, Am Neuen Palais 10, 
  D-14469 Potsdam, Germany\\
  \email{lutz@astro.physik.uni-potsdam.de}
\and Institut f\"ur Philosophie der Universit\"at Bern, L\"anggassstrasse 49a,
   CH-3012 Bern, Switzerland\\
   \email{gerd.grasshoff@philo.unibe.ch}
}

\offprints{nchristlieb@hs.uni-hamburg.de}
\date{Received 10-April-2002; accepted 5-June-2002}
\titlerunning{Automatic spectral classification}
\authorrunning{Christlieb et al.}

\abstract{ We employ classical statistical methods of multivariate
  classification for the exploitation of the stellar content of the
  Hamburg/ESO objective prism survey (HES). In a simulation study we
  investigate the precision of a three-dimensional classification
  ($T_{\mbox{\scriptsize eff}}$, $\log g$, [Fe/H]) achievable in the HES for
  stars in the effective temperature range
  $5200\,\mbox{K}<T_{\mbox{\scriptsize eff}}<6800\,\mbox{K}$, using Bayes
  classification. The accuracy in temperature determination is better than
  400\,K for HES spectra with $S/N>10$ (typically corresponding to
  $B_J<16.5$).  The accuracies in $\log g$ and [Fe/H] are better than
  $0.68$\,dex in the same $S/N$ range. These precisions allow for a very
  efficient selection of metal-poor stars in the HES. We present a minimum
  cost rule for compilation of complete samples of objects of a given class,
  and a rejection rule for identification of corrupted or peculiar spectra.
  The algorithms we present are being used for the identification of other
  interesting objects in the HES data base as well, and they are applicable to
  other existing and future large data sets, such as those to be compiled by
  the DIVA and GAIA missions.  \keywords{Surveys -- Methods: data analysis --
    stars: fundamental parameters -- Galaxy: halo} }

\maketitle

\section{Introduction}

Ever since powerful computers and digital spectra have become available, there
have been efforts to develop algorithms for automatic spectral classification
\citep[for a review on the early works see][]{Kurtz:1984}. The advantages
of automated procedures as compared to manual classification are obvious.
First of all, only a few experts are able to perform accurate manual
classifications, and it was therefore sought to ``freeze'' this expert
knowledge into computer programs. Such programs would allow to obtain
\emph{objective} classifications by \emph{quantitative} criteria, and much
larger data sets could be processed than by manual classification. The latter
issue has become ever more demanding, with upcoming survey missions like
DIVA\footnote{\texttt{http://www.ari.uni-heidelberg.de/diva/}},
NGST\footnote{\texttt{http://ngst.gsfc.nasa.gov/}}, or
GAIA\footnote{\texttt{http://astro.estec.esa.nl/GAIA/}}. With all these
satellites, it is planned to detect millions of objects, or even one billion
objects in the case of GAIA.

{\small
\begin{table*}[htbp]
\caption{\label{performance_comparison} Comparison of automatic spectral classification performances. }
\begin{center}
  \begin{tabular}{lllllllll}\hline\hline
     Method & Type of spectra & $\lambda$ range & Disp. & $S/N$ & Types &
     $\sigma_{\mbox{\scriptsize type}}$ & $\sigma_{\mbox{\scriptsize LC}}$ & Reference\rule{0ex}{2.5ex}\\\hline 
     PCA     & Slit/photoelectric & 3500--4000\,{\AA} & 10\,{\AA}/px & & A0--G0 & 1.16 & 0.85 & W83\rule{0ex}{2.5ex}\\
     Metric dist. & Slit/CCD & 3800--5190\,{\AA} & 67\,{\AA}/mm & & F8--G8 & 0.4 & & LS94\\
     Metric dist. & Slit/CCD & 3500--5100\,{\AA} & 1--2\,{\AA}/px & & B0--F5 & 1.5 & & P94\\
     ANN     & IUE & 1150--3200\,{\AA} & 2\,{\AA}/px & & O3--G5 & 1.11 & & VP95\\
     Metric dist. & IUE & 1150--3200\,{\AA} & 2\,{\AA}/px & & O3--G5 & 1.38 & & VP95\\
     ANN & Slit/Reticon & 5750--8950\,{\AA} & 7\,{\AA}/px & & A0--A9 & $0.42^{\ast}$ & $0.15^{\ast}$ & WTD95\\
     ANN & Slit/Reticon & 5750--8950\,{\AA} & 7\,{\AA}/px & & O4--M6 & $1.26^{\ast}$ & $0.38^{\ast}$ & WTD97\\
     ANN+PCA & Slit/CCD & 3510--6800\,{\AA} & 5\,{\AA}/px & & O--M & 2.34 &      & SGG98\\
     Manual  & Objective prism, widened & 3800--5190\,{\AA} & 108\,{\AA}/mm & $>100$? & B2--M7 & $0.6^{\ast\ast}$
        & $0.25^{\ast\ast}$ & H75--88\\
     Metric dist. & Digitized objective prism & 3800--5190\,{\AA} & 1--3\,{\AA}/px & $>100$? & B & 1.14 & & LS94\\
     ANN   & Digitized objective prism & 3800--5190\,{\AA} & 1--3\,{\AA}/px & $>100$? & B2--M7 & $0.82^{\ast\ast\ast}$
        & & BJIvH98\\
     ANN   & Slit/CCD                  & 3850--4450\,{\AA} & 0.65\,{\AA}/px & $>20$ & F5--K5 & $0.57$--$0.64$ &
        & Setal01\\
     Bayes & Digitized objective prism & 3200--5300\,{\AA} & 7--18\,{\AA}/px & 10--30 & F2-K0 & $<1.6$ & $<0.55$ &
        This work\\\hline\hline\\[-1.5ex]
     \multicolumn{9}{l}{References: W83=\cite{Whitney:1983}; LS94=\cite{LaSala:1994}; P94=\cite{Penprase:1994}; 
                        VP95=\cite{Viera/Ponz:1995};}\\
     \multicolumn{9}{l}{\hspace{4ex}WTD95=\cite{Weaver/Torres-Dodgen:1995}; WTD97=\cite{Weaver/Torres-Dodgen:1997}
                        SGG98=\cite{Singhetal:1998};}\\
     \multicolumn{9}{l}{\hspace{4ex}Houk75--88=\cite{Houk:1975}, \cite{Houk:1978}, \cite{Houk:1982}, \cite{Houk/Smith-Moore:1988};
                        BJIvH98=\cite{Bailer-Jonesetal:1998a};}\\
     \multicolumn{9}{l}{\hspace{4ex}Setal01=\cite{Snideretal:2001}}\\
     \multicolumn{9}{l}{$^{\ast}$ Mean absolute deviation}\\
     \multicolumn{9}{l}{$^{\ast\ast}$ According to \cite{vonHippeletal:1994a}}\\
     \multicolumn{9}{l}{$^{\ast\ast\ast}$ 68\,\% quantile}
   \end{tabular}
\end{center}  
\end{table*}
}

In the last decade, much progress was made in the field of automatic spectral
classification, and it was demonstrated that computers are actually capable of
performing this task \citep[for a recent, comprehensive review
see][]{Bailer-Jones:2001}. Using Kurtz' metric distance approach
\citep{Kurtz:1984}, \cite{LaSala:1994} automatically classified digitized
objective prism spectra from Houk's plates, with good results ($\sigma=1.14$
MK-types). \cite{Penprase:1994} used a similar approach, and applied it to
slit spectra with similar spectral resolution and a slightly larger wavelength
coverage (see Tab. \ref{performance_comparison} for a comparison of the data
used, and results obtained). The spectral type accuracy he reached for B0--F5
stars was a bit worse than that of LaSala; i.e., $\sigma=1.5$ MK-types.
However, as we will see below, it is very difficult to compare the performance
of classification algorithms based on the results published in the literature,
because (a) rarely ever is the signal-to-noise ratio ($S/N$) of the data
documented, and the achievable classification accuracy depends critically on
$S/N$; (b) different wavelength ranges and spectral resolutions were used; and
(c) the algorithms were applied to stars in differing ranges of spectral type.

The influence of the latter on the achievable classification accuracy is
nicely demonstrated by comparing the results of
\cite{Weaver/Torres-Dodgen:1995} with those of
\cite{Weaver/Torres-Dodgen:1997}. In the former paper, the authors report on
supervised automatic classification of stars of spectral type A0--A9 with a
multi-layer artificial neural network (ANN) with one hidden layer, trained
with a back-propagation algorithm. They reached a mean absolute deviation of
0.42 spectral types and 0.15 luminosity classes. In the second paper, the ANN
was applied to stars in the range O4--M6, and the mean absolute deviations
were only 1.26 spectral types and 0.38 luminosity classes.  The results of
Weaver \& Torres-Dodgen have also shown that spectral classification in the
near infrared can be done with the same accuracy as in the ``classical'' MK
spectral range, with spectra of much lower resolution.  The resolution used by
Weaver \& Torres-Dodgen was only 7\,{\AA} per pixel, and their spectral range
5750--8950\,{\AA}. Their results are comparable to that achieved by others at
three times higher spectral resolution in the optical or UV.

To continue with our brief review, in recent years, ANNs have been
successfully used for supervised automatic spectral classification by a couple
of groups. All of them used multilayer back-propagation networks (MBPNs).
\cite{Viera/Ponz:1995} automatically classified spectra of O3--G5 stars
obtained with the International Ultraviolet Explorer (IUE; dispersion 2\,{\AA}
per pixel) with an MBPN. The 1\,$\sigma$ error was 1.11 spectral
types. They found their ANN classification to be superior to a classification
with a metric distance method ($\sigma=1.38$ types). The data used by
\cite{Singhetal:1998} were optical (3500--6800\,{\AA}) slit spectra with a
dispersion of 5\,{\AA} per pixel. They used Principal Component Analysis (PCA)
to pre-process their spectra, and reduce the number of input nodes. They
obtained an accuracy of 2.34 spectral types over the full MK range (O--M).
\cite{Bailer-Jonesetal:1998a} used again Houk's plate material, digitized with
the APM plate scanner, yielding a wavelength range of 3800--5190\,{\AA} and a
dispersion of 1--3\,{\AA} per pixel. Their best ANN configuration classified
these spectra with an error distribution having a 68\,\% quantile of 0.82
types, and the luminosity classification was correct for 95\,\% of the test
sample spectra.

Recently, \cite{Snideretal:2001} used a MBPN for derivation of the
stellar parameters $T_{\mbox{\scriptsize eff}}$, $\log g$ and [Fe/H] from
moderate resolution (0.65\,{\AA} per pixel) spectra. Although their aim is to
assign \emph{continuous} parameter values to each spectrum, while we as well as
the above mentioned authors carried out \emph{discrete} classifications, we
include their work in our review because Snider et al.  applied their
technique to metal-poor stars, which is also the object type we are mainly
concerned with in this paper.  Snider et al. report classification accuracies
of $\sigma_{T_{\mbox{\tiny eff}}}=135$--$150$\,K, $\sigma_{\log
  g}=0.25$--$0.30$\,dex and $\sigma_{\mbox{\tiny [Fe/H]}}=0.15$--$0.20$\,dex.
However, it appears from the upper panel of their Fig. 4 that subgiants and
horizontal branch stars have been excluded from the sample of stars they
studied. A rough graphical analysis of their Fig. 4 reveals that unlike in
real samples of stars emerging e.g. from wide-angle spectroscopic surveys,
which \emph{do} contain subgiants and horizontal-branch stars, their sample
can be classified in $\log g$ with a similar precision by dividing it into two
classes ``by hand'', that is, assigning $\log g=2.5$ to all stars with
$T_{\mbox{\scriptsize eff}}<5000$\,K, and $\log g=4.5$ to all stars with
$T_{\mbox{\scriptsize eff}}>5000$\,K. Furthermore, it is questionable that
there is any feature present in their set of spectra which does allow for a
gravity classification, since they used continuum divided spectra. The Balmer
jump, which is a gravity indicator in cool stars, is therefore removed. In
conclusion, while Snider et al. succeeded in using ANNs for automated
classification in $T_{\mbox{\scriptsize eff}}$ and [Fe/H], it remains to be
demonstrated with a realistic sample that rectified moderate-resolution
spectra indeed contain the information needed for a useful gravity
classification.


%
%

ANN techniques and ``classical'' statistical methods such as Bayes and
minimum cost rule classifications often perform equally well, in terms of e.g.
minimising the total number of misclassifications. In the present work, we
employ statistical methods, because their mathematical properties are
well-studied, and the formulation of classification rules in the framework of
mathematical statistics makes them very transparent.

Before we go into details of the methods we developed (Sect.
\ref{Sect:Autoclass}), we give a brief overview of the Hamburg/ESO Survey
(HES) in Sect. \ref{Sect:HES}, for better readibility. In Sect.
\ref{Sect:Performance} we investigate the classification performance for stars
in the effective temperature range $5200\,\mbox{K}<T_{\mbox{\scriptsize
    eff}}<6800\,\mbox{K}$ achievable in the HES, by a simulation study. We
summarize our conclusions in Sect. \ref{Sect:Conclusions}.

\section{The Hamburg/ESO Survey}\label{Sect:HES}

\begin{figure*}[htbp]
  \begin{center}
    \epsfig{file=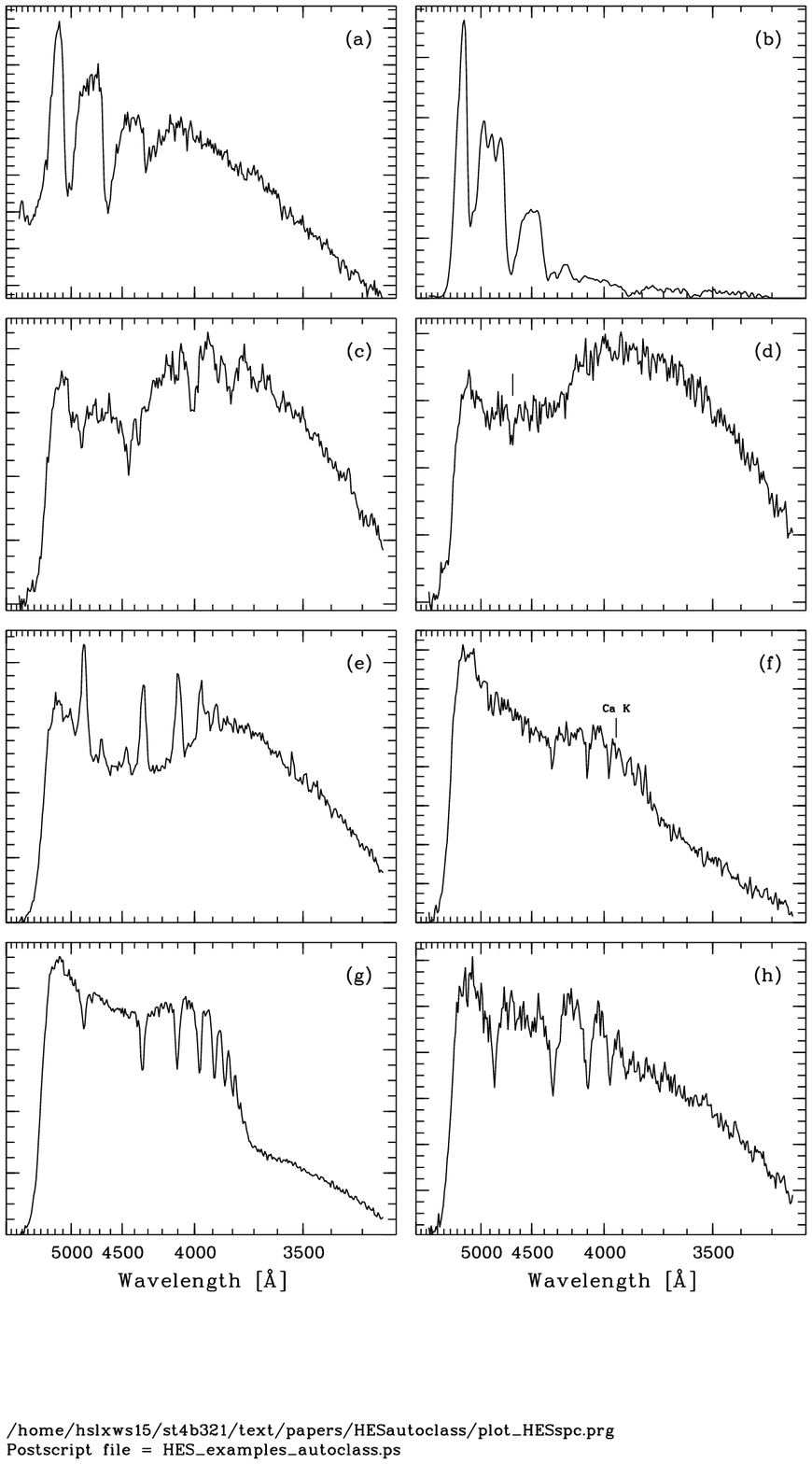, clip=, width=13cm,
      bbllx=81, bblly=160, bburx=437, bbury=727}
  \end{center}
  \caption{\label{Fig:ExampleSpectra} HES example spectra, illustrating the
    large variety of object types that can be identified in that survey.
    Abscissae are wavelength in {\AA}, ordinates are photographic density
    in arbitrary units. Note that wavelength is \emph{de}creasing
    from left to right. The sharp drop at $\sim 5400$\,{\AA} is due to the
    IIIa-J emulsion sensitivity cutoff (``red edge''). Spectra of the
    following object types are shown:
    (a) DQ white dwarf (the red edge of this spectrum is disturbed by an
    overlapping spectrum); (b) cool carbon star; (c) DB white
    dwarf; (d) PG~1159 star (the blend of He~II $\lambda 4686$ and [CIV]
    $\lambda 4660$ is marked); (e) cataclysmic variable star; (f) extremely
    metal-poor star (showing a very weak Ca~K line); (g) FHB/A star;
    (h) cool DA white dwarf. The lower two spectra demonstrate that the strength
    of the Balmer jump can be used as an indicator for the surface gravity
    $\log g$: While it is invisible in the$\log g\sim 8$ white dwarf, the Balmer
    jump is pronounced in the $\log g\sim 2$ horizontal branch star
    }
\end{figure*}

The HES \citep{hespaperI,hespaperIII} is an objective-prism
survey designed to select bright ($12.5 \gtrsim B_J \gtrsim 17.5$) quasars in
the southern extragalactic sky ($\delta<+2.5^\circ$; $|b|\gtrsim 30^\circ$).
It is based on IIIa-J plates taken with the 1\,m ESO Schmidt telescope and its
4$^{\circ}$ prism. The plates were digitized at Hamburger Sternwarte. The HES
spectra cover a wavelength range of $3200\,\mbox{\AA} < \lambda <
5200\,\mbox{\AA}$ and have a seeing-limited spectral resolution of typically
15\,{\AA} at H$\gamma$ and 10\,{\AA} at Ca~II~K 3934\,{\AA}. This resolution
makes it possible to also exploit the stellar content of the survey very
efficiently. For HES example spectra see Fig. \ref{Fig:ExampleSpectra}.

\begin{table*}[htbp]
  \caption{\label{allfeatures} Automatically measured spectral features in the
    HES. The measurement methods of the features \#1--10, \#17--20 and
    \#25--27 have been described in \cite{HESStarsI} and
    \cite{HESStarsII}. The line indices
    {\tt KP}, {\tt GP} and {\tt HP} were proposed by \cite{BeersCaKI}, and the
    definition of {\tt HP} was later refined by \cite{Beersetal:1999}. We
    adapted these indices for the lower resolution of the HES spectra (in
    particular, the positions of some of the continuum bands were changed), and
    calibrated them against stars of Beers et al. present on HES plates. The
    $1\,\sigma$ scatters of these calibrations are $1.22$\,{\AA},
    $1.41$\,{\AA} and $1.61$\,{\AA}, respectively. {\tt balmsum} can be used
    to predict {\tt HP} with an accuracy of $\sigma =1.55$\,{\AA}, and this
    feature is therefore superior to the directly derived {\tt HP}. From
    the half power point distances {\tt dx\_hpp1} and {\tt dx\_hpp2}, $U-B$
    and $B-V$ colors can be derived with accuracies of better than $0.1$\,mag; $c_1$
    can be measured in HES spectra with a precission of $0.15$\,mag
    \citep{HESStarsI}
    }
  \begin{center}
    \begin{tabular}{rlll}\hline\hline
      Number & Name & Description & Measurement method \\\hline
      \#1  & {\tt all5160eqw} & $W_{\lambda}$ of Mg\,{\sc i}~b triplett/TiO $\lambda$\,5168
           & Iterative fit procedure\\
      \#2  & {\tt all4861eqw} & $W_{\lambda}$ of H$\beta$  & Iterative fit procedure\\
      \#3  & {\tt all4388eqw} & $W_{\lambda}$ of Fe {\sc i} $\lambda$\,4383+85 & Iterative fit procedure\\
      \#4  & {\tt all4340eqw} & $W_{\lambda}$ of H$\gamma$  & Iterative fit procedure\\
      \#5  & {\tt all4300eqw} & $W_{\lambda}$ of G-Band & Iterative fit procedure\\
      \#6  & {\tt all4261eqw} & $W_{\lambda}$ of Cr {\sc i} $\lambda$\,4254 + 75 + Fe {\sc i}
        4260 + 72  & Iterative fit procedure\\
      \#7  & {\tt all4227eqw} & $W_{\lambda}$ of Ca {\sc i} $\lambda$\,4227 & Iterative fit procedure\\
      \#8  & {\tt all4102eqw} & $W_{\lambda}$ of H$\delta$ & Iterative fit procedure\\
      \#9  & {\tt all3969eqw} & $W_{\lambda}$ of Ca H + H$\epsilon$ & Iterative fit procedure\\
      \#10 & {\tt all3934eqw} & $W_{\lambda}$ of Ca K & Iterative fit procedure\\
      \#11  & {\tt balmsum}    & {\tt all4861eqw}$+${\tt all4340eqw}$+${\tt
      all4102eqw} & Meta-feature\\
      \#12  & {\tt CaBreak\_sn} & $S/N$ Calcium-break & Template matching\\
      \#13 & {\tt CaBreak\_cont} & Contrast of Calcium-break to continuum
        & Template matching\\
      \#14 & {\tt KP}       & Strength of Ca~K  & Ratio of average pixel values\\
      \#15 & {\tt GP}       & Strength of G band & Ratio of average pixel values\\
      \#16 & {\tt HP}       & Strength of H$\delta$ & Ratio of average pixel values\\
      \#17 & {\tt C2idx1} & Strength of $\mbox{C}_2\;\lambda\,5165$ & Ratio of average pixel values\\
      \#18 & {\tt C2idx2} & Strength of $\mbox{C}_2\;\lambda\,4737$ & Ratio of average pixel values\\
      \#19 & {\tt CNidx2} & Strength of $\mbox{CN}\;\lambda\,4216$ & Ratio of average pixel values\\
      \#20 & {\tt CNidx3} & Strength of $\mbox{CN}\;\lambda\,3883$ & Ratio of average pixel values\\
      \#21 & {\tt klcomp\_1} &  1. continuum shape coefficient & PCA\\
      \#22 & {\tt klcomp\_2} &  2. continuum shape coefficient & PCA\\
      \#23 & {\tt klcomp\_3} &  3. continuum shape coefficient & PCA\\
      \#24 & {\tt klcomp\_4} &  4. continuum shape coefficient & PCA\\
      \#25 & {\tt dx\_hpp1} & Half power point distance 1 & Summing of pixel values\\
      \#26 & {\tt dx\_hpp2} & Half power point distance 2 & Summing of pixel values\\
      \#27 & $c_1$          & Str\"omgren medium band color index & Function of summed pixel values\\\hline\hline
  \end{tabular}
\end{center}
\end{table*}

The goals of automatic spectral classification in the HES are (a)
three-dimensional classification ($T_{\mbox{\scriptsize eff}}$, $\log g$,
[Fe/H]) of the total HES data base currently used for the exploitation of the
stellar content, consisting of $\sim 4$ million spectra, (b) compilation of
complete samples of objects of specific classes, and (c) identification of
peculiar objects. These goals are similar to those emerging in DIVA and GAIA.
Interesting classes of stars that can be found on HES plates include extremely
metal-poor halo field stars \citep{Christlieb/Beers:2000}, field horizontal
branch stars (Christlieb et al., in preparation), carbon stars
\citep{HESStarsII}, and white dwarfs \citep{HESStarsI}. A large data base with
spectra of known type is also very useful for cross-identification with
surveys in other wavelength ranges, such as FAUST \citep[as has been
demonstrated by][]{Broschetal:2000}, the Two Micron All Sky
Survey\footnote{\texttt{http://www.ipac.caltech.edu/2mass/}} (2MASS), or the
Deep Near Infrared
Survey\footnote{\texttt{http://cdsweb.u-strasbg.fr/denis.html}} (DENIS).
Furthermore, the data from these surveys can be used to extend the feature
vectors associated with HES spectra (see below), and improve the automatic
classification in the HES.

\section{Automatic spectral classification}\label{Sect:Autoclass}

In order to achieve our classification aims, we need to construct a
\emph{decision rule} which allows us to assign a spectrum with feature vector
$\vec{x}$ to one of the $n_c$ classes $\Omega_j$, $j=1\dots n_c$, defined in
the specific classification context. That is, we want to carry out a
supervised classification, as opposed to unsupervised classification, where
the aim is to group objects into classes not defined before the classification
process.  Methods of unsupervised classification and their application to HES
spectra are presented in \cite{Hennig/Christlieb:2002}.

\subsection{Feature space}\label{Sect:FeatureSpace}

The HES data base of digital spectra can be represented by feature vectors
$\vec{x}$, consisting of a set of continuous variables $x_i$, i.e.
\begin{equation}
\vec{x} = (x_1,\dots,x_d),
\end{equation}
where $d$ is the number of features used. It is critical for automatic
classification to have a set of reliable features at hand. A wide range of
spectral features is automatically measured in the digitized HES
objective-prism spectra during the data reduction process (cf. Tab.
\ref{allfeatures}): stellar absorption and emission lines, absorption bands,
continuum shape including spectral breaks, bisecting points of spectral
density distribution.  These features are measured in unfiltered HES spectra
with the methods described in \cite{HESStarsI}.

\subsection{Choosing a feature combination}\label{Sect:FeatureCombination}

It is necessary to select a subset of the available features for each
classification problem, and each $S/N$ level, because of several reasons.
\begin{enumerate}
\item[(1)] Blended lines, e.g. H$\varepsilon$+Ca~H, can confuse the
  classification.
\item[(2)] It is advantageous to exclude redundant features from the set of
  features used for classification, since the usage of fewer features
  results in more stable estimates of the parameters of the multivariate
  normal distributions (see Eq. (\ref{multigauss}) below).
\item[(3)] The optimal feature set can vary with $S/N$. For instance, at low
  $S/N$ it can be useful to only use continuum shape parameters and colors
  for classification, because no stellar lines can be detected reliably
  anymore.
\end{enumerate}

The evaluation of the suitability of all $2^d-1$ possible combinations of $d$
available features for a given classification problem is a complex task at
first glance. However, it is usually possible to select a set of appropriate
features by physical considerations alone. E.g., when it is desired to select
metal-poor stars, only those features that are possibly useful as indicators
for $T_{\mbox{\scriptsize eff}}$, $\log g$, [Fe/H], and [C/Fe] need to be
considered.  By means of accuracy considerations and parameter studies,
further features can be rejected. Finally, it is also possible to reduce the
dimensionality of the feature space by \emph{a priori} combining redundant
features, e.g. the equivalent widths of the Balmer lines to a sum of
equivalent widths. The remaining set of features is then evaluated with the
methods described in Sect. \ref{Sect:Evalu}.

\subsection{Learning sample}

For supervised classification, a \emph{learning sample} is needed. For our
purposes, we define a learning sample to be a set of $n_{l}$ objects for which
the feature vectors are known,
\begin{equation}
  \{\vec{x}\} = (\vec{x}_1,\dots,\vec{x}_{n_{l}}),
\end{equation}
and for which it is known to what class they belong. These classes can be
defined, e.g., by grouping a set of objects according to their stellar
parameters (e.g. $T_{\mbox{\scriptsize eff}}$, $\log g$, [Fe/H]), or by
manually assigning classes to a set of spectra by comparison with reference
objects.  With the help of a learning sample, information on the
class-conditional probability densities $p(\vec{x}|\Omega_j)$ can be gained.
$p(\vec{x}|\Omega_j)d\!\vec{x}$ is the probability to observe a feature vector
in the range $\vec{x}\dots \vec{x}+d\!\vec{x}$ in the class $\Omega_j$.  We
inspected the one-dimensional class-conditional probability distributions of
the classes covered by the learning samples used in this work, and
qualitatively found their shapes to agree well with Gaussians. We hence model
$p(\vec{x}|\Omega_j)$ by multivariate normal distributions, i.e.,
\begin{equation}\label{multigauss}
{\small
   p(\vec{x}|\Omega_j) = \frac{1}{(2\pi)^{d/2}\sqrt{|\Sigma_j|}}
   \exp\left\{-\frac{1}{2}\left(\vec{x}-\vec{\mu}_j\right)'
   \Sigma^{-1}_{j}\left(\vec{x}-\vec{\mu}_j\right)\right\},
}
\end{equation}
where $j$ denotes class number, $\vec{\mu}_j$ the mean feature
vector of class $\Omega_j$, and $\Sigma_j$ the covariance matrix of class
$\Omega_j$.

\subsection{Decision rules}\label{Sect:DecisionRules}

A central issue in automatic classification is the construction of a
decision rule which is optimal for the given classification problem. In the
HES, we use three decision rules: the Bayes rule, a minimum cost rule, and
a rejection rule. 

\subsubsection{Bayes' rule}

Classification with Bayes' rule minimizes the total number of
misclassifications, if the true distribution of class-conditional
probabilities $p(\vec{x}|\Omega_i)$ is used \citep{Hand:1981,Anderson:1984}.
Using Bayes' theorem,
\begin{equation}
  P(\Omega_i|\vec{x})=\frac{P(\Omega_i)p(\vec{x}|\Omega_i)}
  {\sum\limits_{\forall i}P(\Omega_i)p(\vec{x}|\Omega_i)},
\end{equation}
posterior probabilities $P(\Omega_i|\vec{x})$ can be calculated. A spectrum of
unknown class, with given feature vector $\vec{x}$, can then be classified
using Bayes' rule:
\begin{description}
\item[Bayes' rule:] \emph{Assign a spectrum with feature vector $\vec{x}$ to
    the class with the highest posterior probability $P(\Omega_i|\vec{x})$.}
\end{description}

\subsubsection{Minimum cost rule}

In most of the classification problems arising in the HES it is desired to
gather a sample of objects of a specific class, or a specific set of
classes. In these cases, Bayes' rule is not appropriate, because we do not
want to minimize the total number of misclassifications, but the
misclassifications between the desired class(es) of objects, and the remaining
classes. Suppose we have three classes, A-, F-, and G-type stars, and we want
to gather a complete sample of A-type stars. Then only misclassifications
between A-type stars and F- and G-type stars (and vice versa) are of interest.
More specifically, misclassifications of A-type stars to F- and G-type stars
(leading to incompleteness) are least desirable when a complete sample shall
be gathered, and erroneous classification of F- and G-type stars as A-type
stars (resulting in sample contamination) can be accepted at a moderate rate.
Misclassifications between F- and G-type stars can be totally ignored, because
the target object type is not involved.

Classification aims like this can be realized by using a minimum cost rule.
Cost factors $r_{hk}$, with
\begin{equation}\label{Def_Verlustfaktoren}
  0 \le r_{hk} \le 1; \qquad h = 1,\dots,n_c;\quad k = 1,\dots,n_c.
\end{equation}
allow to assign relative weights to individual types of
misclassifications.  The cost factor $r_{hk}$ is the relative weight of a
misclassification from class $\Omega_h$ to class $\Omega_k$.

Suppose we have an object of unknown class, with feature vector $\vec{x}$. We
ask how large the cost is if it belongs to class $\Omega_h$, and would be assigned
to class $\Omega_k$, $h\not= k$. The cost $C_{h\to k}(\vec{x})$ is:
\begin{eqnarray}
C_{h\to k}(\vec{x}) &=& r_{hk}P(\Omega_h|\vec{x})\nonumber\\
        &=& r_{hk}\frac{P(\Omega_h)\, p(\vec{x}|\Omega_h)}{
             \sum\limits_{i=1}^{m}P(\Omega_i)\,p(\vec{x}|\Omega_i)}\nonumber\\
        &=& r_{hk}\frac{a_h p_h(\vec{x})}{
             \sum\limits_{i=1}^{m}a_i p_i(\vec{x})}.\label{Vhk}
\end{eqnarray}
In the last step we have used the abbreviations $P(\Omega_h)=a_h$ and
$p(\vec{x}|\Omega_h)=p_h(\vec{x})$. We do not know to which of the possible
classes $\Omega_h$, $h = 1,\dots,n_c$, the object actually belongs. Therefore,
we estimate the expected cost $C_k(\vec{x})$ for assigning an object with
feature vector $\vec{x}$ to the class $\Omega_k$ by computing the following
sum of costs:
\begin{eqnarray}
  C_k(\vec{x}) &=& \sum\limits_{h=1\atop h\not=k}^{m}
                   C_{h\to k}(\vec{x})\nonumber\\
               &=& \sum\limits_{h=1\atop h\not=k}^{m} r_{hk}\frac{a_h
                 p_h(\vec{x})}{\sum_{i=1}^{m}a_i p_i(\vec{x})}\label{V_k}
\end{eqnarray}

\noindent Now we can formulate the minimum cost rule, which minimizes
the total cost \citep{Hand:1981}.
\begin{description}
\item[Minimum Cost Rule:] {\em Assign an object with feature vector $\vec{x}$
    to the class $\Omega_k$ with the lowest expected cost $C_k(\vec{x})$.}
\end{description}

If the cost factors are chosen such that $r_{hk}\equiv\delta_{hk}$,
the minimum cost rule classification is identical to classification
using Bayes' rule. In this case the cost for assigning the
class $\Omega_k$ to a spectrum with feature vector $\vec{x}$ is the
probability that the object belongs to one of the other classes
$h\not= k$. This follows immediately from Eq. (\ref{V_k}). If
$r_{hk}\not=\delta_{hk}$, the total number of misclassifications is
\emph{not} minimized, so that the quality of a minimum cost rule
classification has to be evaluated by other criteria.

For any given classification aim, one can divide the cost factors to
be chosen into three sets:
\begin{description}
\item[\texttt{t2o:}] Cost factor for misclassification of an object of the
    {\bf t}arget class (`\texttt{t}') to (`\texttt{2}') one of the {\bf
      o}ther classes (`\texttt{o}').
\item[\texttt{o2t:}] Cost factor for contamination of the target class.
\item[\texttt{o2o:}] Cost factor for misclassification between other classes.
\end{description}
Since sample completeness and contamination are interdependent, in practice
only the relative value \texttt{t2o}/\texttt{o2t} has to be adjusted.  For
this purpose, the classification results as a function of
\texttt{t2o}/\texttt{o2t} are evaluated. The expected error rates, estimated
e.g. with the ``leaving one out'' method (see Sect. \ref{Sect:Evalu} below),
tell which level of completeness and sample contamination will be achieved.
\cite{ncADASSVII} presented a software tool for a convenient choice of cost
factors. 

\subsubsection{Rejection rule}\label{Sect:RejectRule}

Non-mathematically speaking, Bayes' rule assigns the class with the highest
relative resemblance to each spectrum to be classified.  However, it is
ignorant of the absolute resemblance: A spectrum with feature vector $\vec{x}$
may be assigned to a class with very low posterior probability
$p(\Omega_i|\vec{x})$, if $p(\Omega_i|\vec{x})$ is even lower for all other
classes. This means that a class is assigned to all spectra, even to ``garbage
spectra'' which are disturbed, for instance, by plate artifacts. Therefore, it
is useful to apply a rejection rule in addition to either the Bayes rule or
the minimum cost rule. The rejection rule can also be used ``stand alone'' for
the identification of peculiar objects, e.g., quasars.

\begin{description}
\item[Rejection rule:] {\em Reject an object from classification to class
  $\Omega_i$, if $A(\Omega_i;\vec{x}) > \beta\,$}.
\end{description}

The parameter $\beta$ is a threshold to be chosen, and the parameter $A$ is
the {\em atypicality index\/} suggested by \cite{Aitchisonetal:1977},
\begin{equation}
  A(\Omega_i,\vec{x}) = \Gamma\left\{\frac{d}{2};\frac{1}{2}
      \left(\vec{x}-\vec{\mu}_i\right)'
      \Sigma^{-1}_{i}\left(\vec{x}-\vec{\mu}_i\right)\right\},
\end{equation}
where $\Gamma(a;x)$ is the incomplete gamma function and $d$ the number of
features used for classification. Use of the above rejection criterion is
identical to performing a $\chi^2$ test of the null hypothesis $H_0$ that an
object with feature vector $\vec{x}$ belongs to class $\Omega_i$ at
significance level $1-\beta$, against the alternative hypothesis $H_1$ that it
does belong to class $\Omega_i$. We reject the null hypothesis if its
significance level is low, i.e., if it is very unlikely that a feature vector
$\vec{x}$ is observed for class $\Omega_i$, given the multivariate normal
distributions (\ref{multigauss}) are the true distributions of the
class-conditional probabilities $p(\vec{x}|\Omega_i)$.

\section{Classification performance}\label{Sect:Performance}

\begin{figure*}[htbp]
  \begin{center}
    \leavevmode
    \epsfig{file=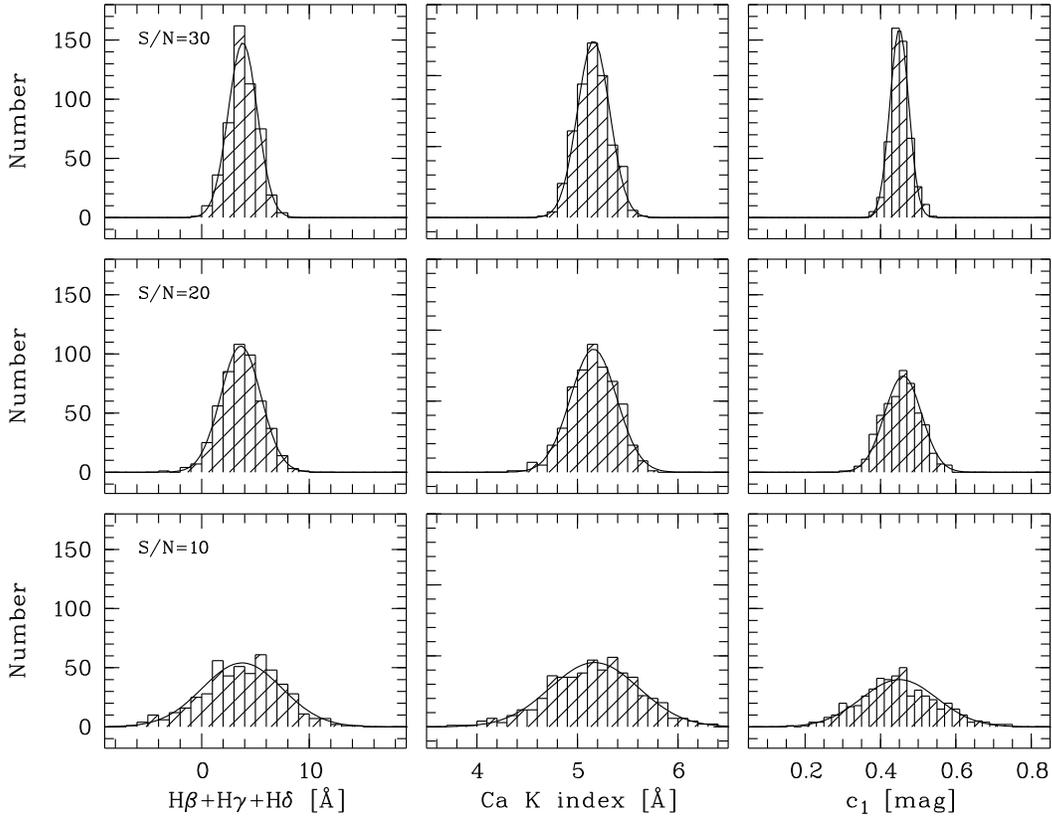, clip=, width=14cm,
      bbllx=47, bblly=225, bburx=492, bbury=573}
  \end{center}
  \caption{\label{HESpardistrib} Distribution of the three features used
    for classification of main sequence turnoff stars in one learning
    sample class, and for different signal-to-noise ratios.}
\end{figure*}

In a first application of automatic spectral classification in the HES, we
selected candidates for extremely metal-poor halo field stars.
\cite{Christlieb:2000} and \cite{Christlieb/Beers:2000} have shown that with
this method, a very efficient selection of metal-poor stars is feasible.
80\,\% of an investigated sample of 56 highest priority metal-poor candidates
were shown by medium-resolution follow-up spectroscopy to have metallicities
below $-2.0$\,dex, and results based on a larger sample of stars, including
also fainter and lower priority candidates, indicate that the overall
efficiency for the selection of stars with $\mbox{[Fe/H]}<-2.0$ is $\sim
60$\,\% in the HES (Christlieb et al., in preparation). This is the most
efficient selection of metal-poor stars ever obtained in a wide-angle survey
for such stars. In this paper we focus on results of a systematic
investigation of the classification performance for stars in the effective
temperature range $5200\,\mbox{K}<T_{\mbox{\scriptsize eff}}<6800\,\mbox{K}$
achievable in the HES, by means of a simulation study.

\subsection{Evaluation of classification rules}\label{Sect:Evalu}

Classification rules can be evaluated by the number of expected
misclassifications (in the case of Bayes' rule), or by the total expected cost
(in the case of the minimum cost rule). The three most important methods to
estimate these numbers are \citep{Deichsel/Trampisch:1985}:
\begin{enumerate}
\item[(1)] Re-substitution 
\item[(2)] ``Hold out'' method
\item[(3)] ``Leaving one out'' method.
\end{enumerate}

Re-substitution means that one uses the learning sample also as test
sample. The drawback of this method is that one underestimates the
number of expected misclassifications, because a classification rule derived
with the help of a finite learning sample is always adapted to the individual
composition of the learning sample. Therefore, the estimation of the expected
number of misclassifications is biased \citep{Deichsel/Trampisch:1985}.

An improvement in this respect is gained when the ``hold out'' method is used.
Here one randomly divides the learning sample disjointly into a new, smaller
learning sample, and a test sample. Since the learning sample and test sample
are completely independent in this case, an unbiased estimate of the expected
error rates is possible \citep{Deichsel/Trampisch:1985}. However, the drawback
is that one needs a large enough learning sample. When modeling the
class-conditional probabilities with multivariate normal distributions,
the learning sample size has to be large enough to ensure a robust estimation
of the parameters of the distributions. 

The problem of learning sample size can be circumvented by using the ``leaving
one out'' method. Suppose we have a learning sample of size $n_{l}$. We
exclude object $i$ from the learning sample, and construct the classification
rule using the $n_{l}-1$ remaining objects. Object $i$ is then classified
with this classification rule. This procedure is repeated $n_{l}$ times, so
that each object of the learning sample is excluded once, and used as test
sample.  By adding up the numbers of misclassifications obtained in each step,
one gets an unbiased estimate of the expected error rate
\citep{Deichsel/Trampisch:1985}. The only drawback of this method is that it
consumes a lot more computing time than the previously mentioned methods,
since $n_{l}$ classification rules have to be constructed. However, the
computing time increases only linearly with learning sample size
$n_{l}$, so that the usage of the ``leaving one out'' method was feasible for
all HES learning samples used so far (the largest learning sample used
had $n_{l}=165\,000$).

\subsection{Simulation study on the classification performance in the HES}

For our simulation study, we employed a grid of model
spectra converted to objective prism spectra with the methods
described in \cite{HESStarsI}. The grid covers the following stellar
parameter range:
\begin{eqnarray*}
  T_{\mbox{\scriptsize eff}} &=& 5200(200)6800\,\mbox{K}\\
  \log g &=& 2.2(0.8)4.6\\
  \left[\mbox{Fe/H}\right] &=& -0.3,-0.9,-1.5(0.3)-3.6
\end{eqnarray*}
The values in brackets refer to grid point distances. The grid defines 360
classes. Since it is one of the aims of our simulation study to
  investigate how the classification accuracy changes with $S/N$, we need to
  simulate spectra of different $S/N$. For this, we added Gaussian noise to
the grid of simulated spectra so that spectra with $S/N=5(5)30$ resulted.
The adequatness of this noise model for HES spectra has been
  demonstrated by \cite{HESStarsI}. It is necessary to produce learning
samples for different $S/N$ levels because the width of the class-conditional
probability distributions change with $S/N$ (see Fig. \ref{HESpardistrib}).
We then performed a Bayes classification, using the three features $c_1$,
\texttt{balmsum} (the sum of the equivalent widths of H$\beta$, H$\gamma$ and
H$\delta$), and the Ca~K index {\tt KP}. This feature set was found to be best
suitable for the desired three-dimensional classification in parameter
studies, and by systematic evaluation of the classification performance of
different feature combinations.

\begin{figure}[htbp]
  \begin{center}
  \leavevmode
  \epsfig{file=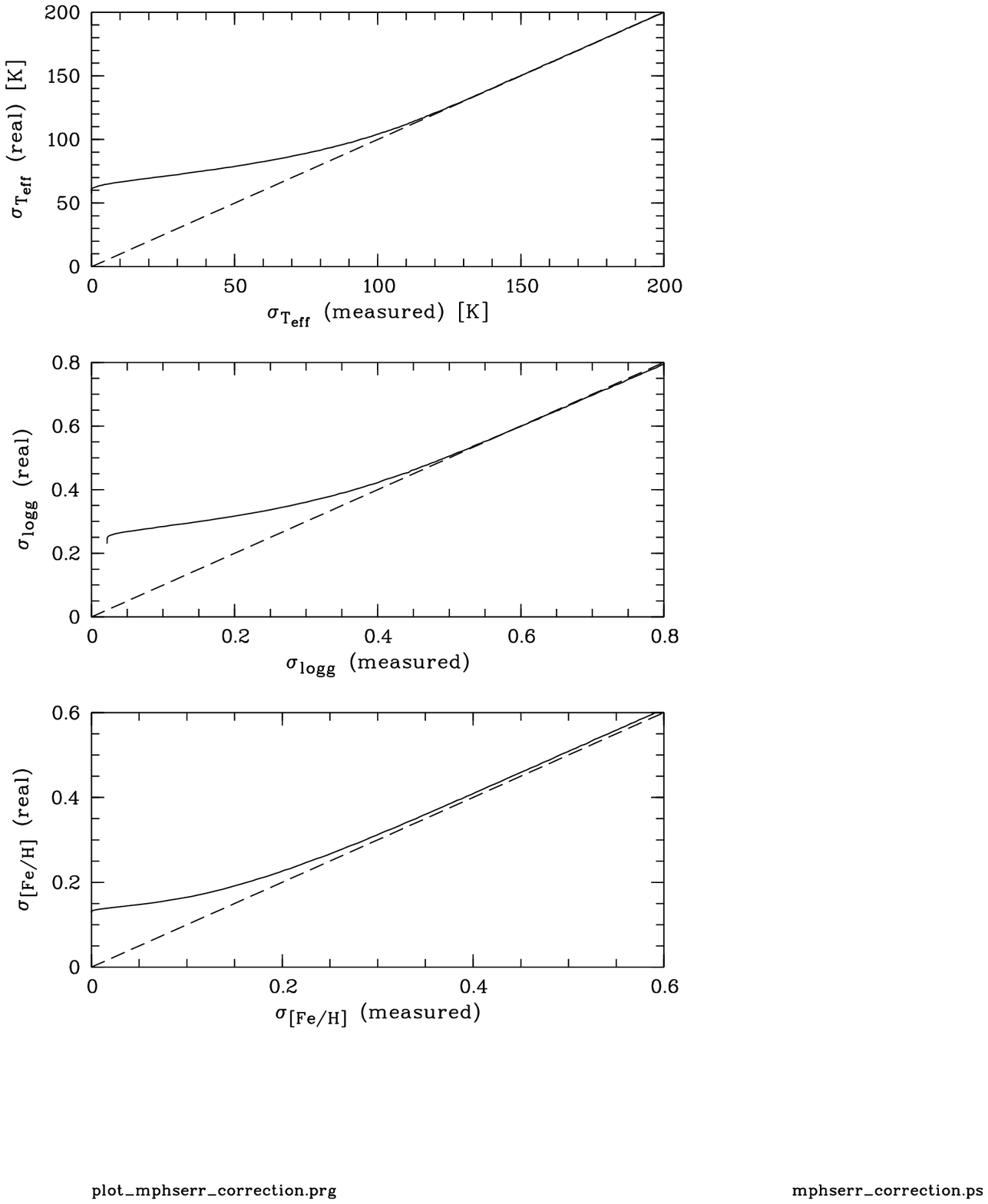, clip=, width=8.8cm,
     bbllx=100, bblly=225, bburx=407, bbury=686}
  \end{center}
  \caption{\label{Fig:error_correction} Corrections applied to the
  classification errors measured in the simulation study. For explanation see
  text}
\end{figure}

For each spectral class 500 simulated spectra were computed, which is a large
enough number to randomly subdivide the grid into a learning sample and an
independent test sample. To obtain a realistic estimate of the classification
performance, two effects have to be taken into account:
\begin{description}
\item[Undersampling of the error distribution:] In our simulation study, we use
  a grid of stellar parameters. Therefore, if a classification error
  is smaller than half of the grid point distance, an error of zero is measured.
  Therefore, the classification error is systematically underestimated in
  these cases. 
\item[Discretization error:] Real samples of stars have a continuous distribution
  of stellar parameters. These parameters will be mapped to our discrete
  grid. This results in classification errors for stars having stellar
  parameters lying between two grid points. 
\end{description}
We have taken these effects into account by applying (upward) corrections to
the classification errors measured on our model spectra grid (see Fig.
\ref{Fig:error_correction}). For the estimation of the corrections for error
distribution undersampling, Gaussian random errors were added to the grid
point parameters to simulate classification errors, and the measured
errors were compared with the errors known from the chosen $\sigma$ of the
Gaussian distribution. The discretization error correction to be applied was
derived by mapping continuously distributed stellar parameters to our grid,
and computing the mean difference between real parameters and the parameters
detected by the grid.

In the stellar parameter range we explored so far, the corrected accuracy in
effective temperature classification is better than 400\,K for spectra with
$S/N>10$, which typically corresponds to $B_J<16.5$ \citep[see][]{HESStarsI}.
The accuracies in $\log g$ and [Fe/H] are better than $0.68$\,dex for the same
magnitude range. Note that the accuracy in [Fe/H] strongly depends on [Fe/H]
itself, since the Ca~K line, used as metallicity indicator, is not detectable
in the spectra of the lowest metallicity turnoff stars at the spectral
resolution of the HES. Therefore, a metallicity classification is not possible
in that part of the stellar parameter space, resulting in a larger average
classification error.

\begin{figure}[htbp]
  \begin{center}
  \leavevmode
  \epsfig{file=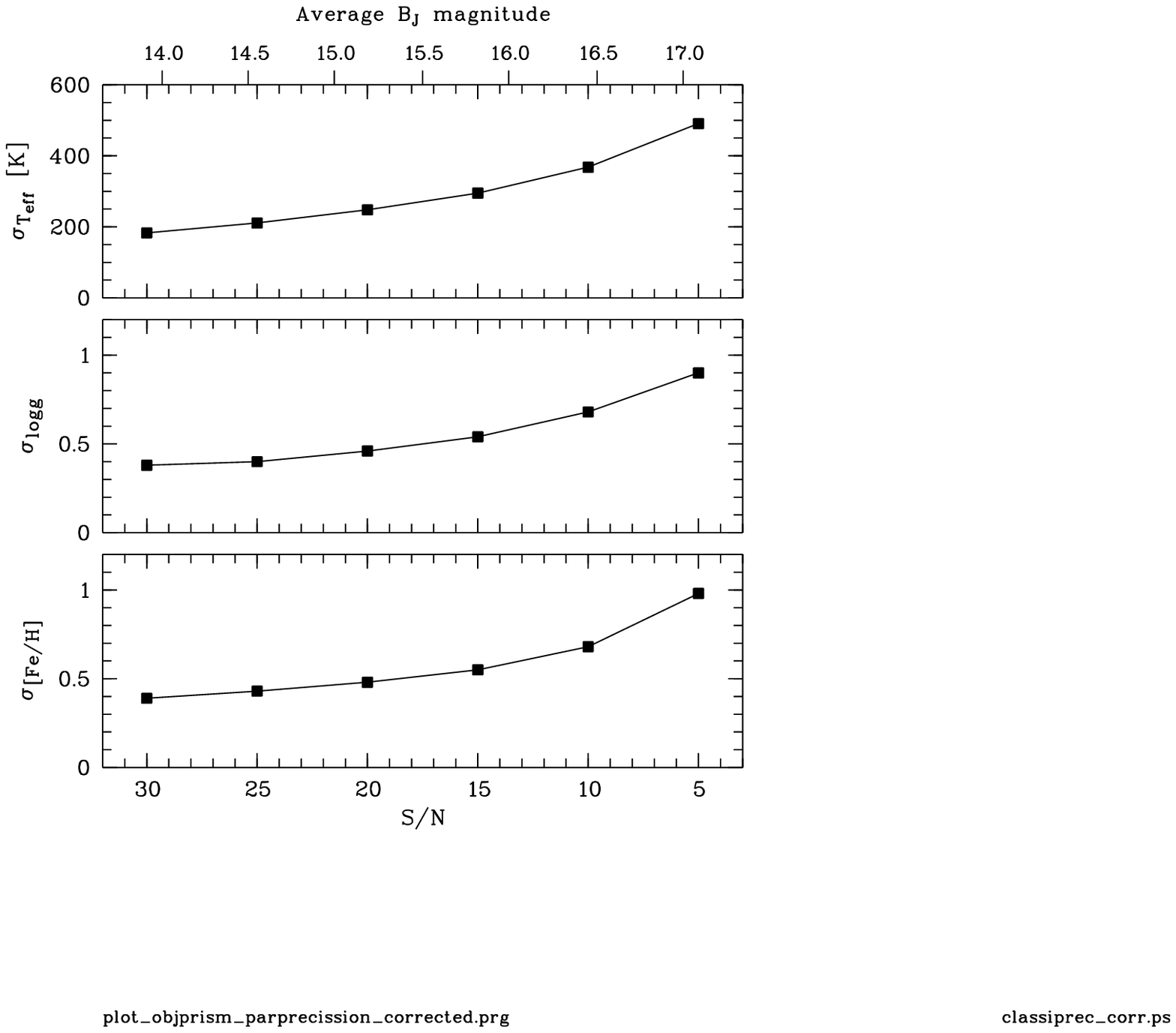, clip=, width=8.8cm,
     bbllx=72, bblly=354, bburx=370, bbury=685}
  \end{center}
  \caption{\label{HESclassiprec} Classification precision for stars in the
    effective temperature range $5200\,\mbox{K}<T_{\mbox{\scriptsize
        eff}}<6800\,\mbox{K}$ in the HES as a function of $S/N$, as obtained
    with Bayes classification in our simulation study}
\end{figure}

As a plausibility check we compared our results with the classification
accuracies we would expect from simple, one-dimensional parameterization
approaches using $B-V$, $c_1$ and {\tt KP} as temperature, gravity and
metallicity indicators, respectively. In the effective temperature range
$5200\,\mbox{K}<T_{\mbox{\scriptsize eff}}<6800\,\mbox{K}$, $\Delta
(B-V)/\Delta T_{\mbox{\scriptsize eff}}\sim 0.028\,\mbox{mag}/100\,\mbox{K}$
\citep{Lang:1992}. The average accuracy of the HES $B-V$ calibration in the
temperature range under consideration, averaged over the full magnitude range
covered by the HES, is $\sigma_{B-V}=0.07$\,mag \citep{HESStarsI}, so that an
average temperature classification accuracy of $\sigma_{T_{\mbox{\tiny
      eff}}}=260$\,K is expected.  This is consistent with classification
errors of 200--420\,K in the magnitude range $14.0<B_J<17.5$. $ \Delta
c_1/\Delta \log g\sim 0.1$--$0.3\,\mbox{mag}/\mbox{dex}$ in the effective
temperature range under consideration \citep{Lang:1992}. The average accuracy
of the HES $c_1$ calibration is $\sigma_{c_1}=0.15$\,mag \citep{HESStarsI}, so
that we expect a gravity classification precision of $\sigma_{\log
  g}=0.5$--1.5\,dex. This is consistent with $\sigma_{\log g}<0.68$ measured
in our simulation. Finally, from Fig. 4 of \cite{BeersCaKI} one can read that
at $B-V=0.5$, the difference in the Ca~K index {\tt KP} between a star of
$\mbox{[Fe/H]}=-2.0$ and a star of $\mbox{[Fe/H]}=-3.0$ is $2.7$\,{\AA} and
$4.7$\,{\AA} for dwarfs and giants, respectively. Considering the fact that
$\sigma_{\mbox{\scriptsize \tt KP}}=1.22$\,{\AA} in the HES, it is not
surprising that classification precisions as high as $0.4$\,dex can be
achieved for the brightest stars in the HES (see Fig. \ref{HESclassiprec}).

\section{Discussion and conclusions}\label{Sect:Conclusions}

We have demonstrated that automatic spectral classification of turnoff stars,
using ``classical'' statistical methods, is feasible in the HES with high
accuracy. Our results suggest that it might be possible to determine the
metallicity distribution function (MDF) of the galactic halo directly from a large
sample of HES spectra. The MDF is an important constraint for models of
galactic chemical evolution \citep[see, e.g.,][]{Ikuta/Arimoto:1999,Oey:2000}.

The described methods are currently being applied to the large HES data base
of digital spectra, in order to select interesting stellar objects in an
automated fashion, and fully exploit the large scientific potential present in
our data base.

Our algorithms can easily be adapted for automatic classification of other
large data sets, e.g. those to be compiled by the DIVA and GAIA missions.

\begin{acknowledgements}
  We thank the other members of the ``methods of scientific discovery'' group
  -- A. Nelke, and A. Schlemminger -- for their valuable contributions.  The
  grid of synthetic spectra was kindly provided by J. Reetz. Discussions with
  C. Hennig on mathematical aspects of our work are acknowledged. We are
  grateful to C. Bailer-Jones for comments on an earlier version of this paper.
  This work was supported by Deutsche Forschungsgemeinschaft under grants
  Re~353/40 and Gr~968/3.
\end{acknowledgements}

\nocite{Corballyetal:1994}
\bibliography{classification,datanaly,HES,mphs,ncpublications,ncastro,statistics}
\bibliographystyle{apj}

\end{document}